\documentclass[mathleft]{an}
\usepackage{graphicx}
\usepackage{times}
\begin{document}

% The following seven commands are intended for editorial usage and should be ignored by
% the author(s).
\Pagespan{789}{}% Document's page range.
% If second parameter is left empty, the last page is computed automatically.
\Yearpublication{2006}%
\Yearsubmission{2005}%
\Month{11}%
\Volume{999}%
\Issue{88}%
% \DOI{This.is/not.aDOI}%

\title{The Effects of Superhigh \\Magnetic Fields on  Equations of States of Neutron Stars}

\author{Z.F. Gao\inst{1,2}\fnmsep\thanks{Corresponding author:
  \email{zhifugao@xao.ac.cn}\newline}
%Example
%for footnote, note the usage of the \texttt{fnmsep}
%command as separator between institute number and footnote mark}
  \and N. Wang\inst{1}\and Y. Xu\inst{3}\and H. Shan\inst{1}\and X.-D. Li\inst{4}\fnmsep\thanks{Corresponding author:
  \email{lixd@nju.edu.cn}\newline}
}
\titlerunning{Magnetic effects on NS EoSs}
\authorrunning{Gao et al.}
\institute{Xinjiang Astronomical Observatory, Chinese Academy of Sciences, Urumqi, Xinjiang, 830011, China
\and Key Laboratory of Radio Astronomy, Chinese Academy of Sciences, West Beijing Road, Nanjing, Jiangsu, 210008, China
\and Changchun Observatory, National Astronomical Observatories, Chinese Academy of Sciences, Changchun, 130117, China
\and Shchool of Astronomy and Space Science, Nanjing University, Nanjing, Jiangshu, 210046 China}

\received{30 May 2005}
\accepted{11 Nov 2005}
\publonline{later}


\abstract{By introducing Dirac $\delta$-function in superhigh magnetic field,
we deduce a general formula for pressure of
degenerate and relativistic electrons, $P_{e}$, which is suitable for
superhigh magnetic fields, discuss the quantization of Landau levels of
electrons, and consider the quantum electrodynamic(QED) effects on the
equations of states (EOSs) for different matter systems. The main
conclusions are as follows: the stronger the magnetic field strength, the
higher the electron pressure becomes; compared with a common radio pulsar,
a magnetar could be a more compact oblate spheroid-like deformed neutron star
due to the anisotropic total pressure; and an increase in the maximum mass of a
magnetar is expected because of the positive contribution of the magnetic
field energy to the EOS of the star. Since this is an original work in which some
uncertainties could exist, to further modify and perfect our theory model
should be considered in our future studies.}

\maketitle

\section{Introduction}
Pulsars are among the most mysterious objects in the universe
that provide natural laboratory for investigating the nature of matter
under extreme conditions, and are universally recognized as normal neutron
stars (NSs), but sometimes have been
argued to be quark stars (Du et al. 2009, Lai et al. 2013, Xu et al. 2013). The equation of
state (EoS) of matter under exotic conditions is an important tool for
understanding of the nuclear force and for astrophysical applications. The
Fermi energy of relativistic electrons $E_{\rm F}(e)$ is one of most
important and indispensable physical parameters in EoS, and affects
direct weak-interaction processes including modified URCA reactions,
electron capture (e.g., Gao et~al.~2011a, 2011b, 2011c, 2011d, 2012a, 2012b; Liu 2012, 2013,
2014, 2015; Du et al. 2014). These influences will change intrinsic EoS, interior
structure and heat evolution, and even affect the whole properties of the star.

As we know, for degenerate and relativistic electrons in $\beta-$equilibrium,
the distribution function $f(E_{e})$ obeys Fermi-Dirac statistics:
$f(E_{e})=1/(Exp((E_{e}-\mu_{e})/kT)+1)$, $k$ represents Boltzmann's
constant, and $\mu_{e}$ is the electron chemical potential. If $T\rightarrow 0$,
$\mu_{e}$ is also called the electron Fermi
energy, $E_{\rm F}(e)$, which presents the energy of highest occupied states for electrons.
The electron Fermi energy $E_{\rm F}(e)$ has the simple form
\begin{equation}
\label{1}
E_{\rm F}(e) = \left(p^{2}_{\rm F}(e)c^{2}+ m^{2}_{e}c^{4}\right)^{1/2}~~,~~%
\end{equation}
with $p_{\rm F}(e)$ being the electron Fermi momentum.

In the context of general relativity principle, the matter density is defined as:
$\rho= \varepsilon/c^{2}$, $\varepsilon$ is the total energy density, including the
rest-mass energies of particles. Using the basic thermodynamics, we obtain the
relation of the total matter pressure $P$ and matter density $\rho$ in a common NS,
\begin{eqnarray}
\label{2}
&&P(n_{B})= n_{B}^{2}\frac{d(\varepsilon/n_{B})}{d n_{B}}~, \nonumber\\
&&\rho(n_{B})= \varepsilon(n_{B})/c^{2}, ~~\Rightarrow P= P(\rho)~. % 2
\end{eqnarray}
From the above equation, it is obvious
that $P$ solely depends on $\rho$. Theoretically, we can obtain the value of $E_{\rm F}(e)$ by
solving EOS in a specific matter model. The pressure of degenerate and relativistic electrons, $P_e$, is another
important and indispensable physical parameter in EoSs of a NS. $P_e$ is one of important dynamical pressures against a NS's
gravitational collapse, and affects the structures and properties of the star, substantially.

Thompson and Duncan (1996) predicted that superhigh magnetic fields (MFs) could
exist in the interiors of magnetars with a typical surface dipolar MF, $B\sim
10^{14}$ to $10^{15}$~G (Thompson \& Duncan 1996). Superhigh MFs have effects on EoSs of a NS, as well as on
its spin-down evolution (e.g., Gao et al. 2014, 2015). Recently, Franzon et al. (2015) studid the effects of strong MFs on hybrid stars by
  using a full general-relativity approach, and pointed that the MF could cause the
  stellar central density to be reduced, inducing major changes in the populated degrees
  of freedom and, potentially, converting a hybrid star into a hadronic star.
In accordance with the popular point of view, the stronger the MF strength, the lower
the electron pressure becomes. With respect to this view,
we cannot directly verify it by experiment in actual existence, owing to lack
of such high MFs on the earth.
After a careful check, we found that popular methods of calculating $E_{F}(e)$ the electron
Fermi energy are contradictory to the quantization of electron Landau levels.
In an extremely strong MF, the Landau column becomes a very long narrow cylinder along MF,
If we consider Dirac $\delta$-function in superhigh MFs, all the results should be re-considered.

In Sec. 2, we deduce an equation of $P_{e}$ in superhigh MFs; in Sec. 3, we consider QED effects
on EOSs of NS matter, and discuss an anisotropy of the total pressure;  In Sec. 4 we
discuss our future work on improving our model, and present conclusions in Sec. 5.
\section{Deduction of the pressure of electrons in superhigh MFs}
The relativistic Dirac-Equation for the electrons in a uniform
external magnetic field along the $z-$axis gives the electron energy level
\begin{equation}
\label{3}
E_{e}= [m_{e}^{2}c^{4}(1+ \nu \frac{2B}{B_{\rm cr}})+ p^{2}
_{z}c^{2}]^{\frac{1}{2}}~~, ~
\end{equation}
where $\nu=n+\frac{1}{2}+
\sigma$ is the quantum number, $n$ the Landau level number, $\sigma
=\pm \frac{1}{2}$ the spin quantum number (Canuto \& Ventura 1977), and $p_{z}$ is the
$z$-component of electron momentum, and may be treated as a
continuous function.  Combining $B_{\rm cr}= m^{2}_{e}c^{3}/{e}\hbar$
with $\mu_{e}^{'}=e\hbar/2m_{e}c$ gives
\begin{eqnarray}
   \label{4}
  E_{e}^{2}&=&m_{e}^{2}c^{4} + p_{z}^{2}c^{2}
 + 2\nu 2m_{e}c^{2}\mu_{e}^{'}B \nonumber\\
 &&= m_{e}^{2}c^{4} + p_{z}^{2}c^{2}
 + p_{\bot}^{2}c^{2}~~,
  \end{eqnarray}
where $\mu_{e}^{'}$ is the magnetic moment of an electron, and $p_{\bot}=m_{e}c(2\nu B^{*})
^{\frac{1}{2}})$.
The maximum electron Landau level number $n_{max}$ is uniquely determined
by the condition $[p_{\rm F}(z)c]^{2}\geq 0$ (Lai \& Shapiro 1991, Gao et al. 2013), where
$p_{\rm F}(z)$ is the Fermi momentum along the $z-$axis.  The
expression for $\nu_{max}$ can be expressed as
\begin{eqnarray}
\label{5}
 && \nu_{max}(\sigma=-\frac{1}{2})\nonumber  \\
 &&=Int[\frac{1}{2B^{*}}[(\frac{E_{\rm F}(e)}{m_{e}c^{2}})
 ^{2}-1 -(\frac{p_{z}}{m_{e}c})^{2}]+\frac{1}{2}-\frac{1}{2}]\nonumber  \\
 &&= Int[\frac{1}{2B^{*}}[(\frac{E_{\rm F}(e)}{m_{e}c^{2}})
 ^{2}-1-(\frac{p_{z}}{m_{e}c})^{2}]]~,\nonumber\\
&& \nu_{max}(\sigma=\frac{1}{2})\nonumber\\
&&=Int[\frac{1}{2B^{*}}[(\frac{E_{\rm F}(e)}{m_{e}c^{2}})
 ^{2}-1-(\frac{p_{z}}{m_{\rm e}c})^{2}]-1+\frac{1}{2}+~\frac{1}{2}]\nonumber\\
 &&=Int[\frac{1}{2B^{*}}[(\frac{E_{\rm F}(e)}{m_{e}c^{2}})
 ^{2}-1-(\frac{p_{z}}{m_{e}c})^{2}]]~.~~
\end{eqnarray}
According to the definition of $E_{\rm F}(e)$ in Eq.(1), we obtain
$E_{\rm F}(e) \equiv p_{\rm F}(e)c$ if electrons are super-relativistic
($E_{\rm F}(e)\gg m_{e}c^{2}$). In the presence of a superhigh MF,
$B\gg B_{\rm cr}$, $E_{\rm F}(e) \gg m_{e}c^{2}$), we have
\begin{eqnarray}
\label{6}
&&\nu_{max}^{'}(\sigma=-\frac{1}{2})=\nu_{max}^{'}(\sigma= \frac{1}{2})\nonumber\\
&&\simeq Int[\frac{1}{2B^{*}}[(\frac{E_{\rm F}(e)}{m_{e}c^{2}})^{2}]~.
\end{eqnarray}
The maximum of $p_{\bot}$ for electrons in a superhigh MF is
\begin{equation}
\label{7}
p_{\bot}^{2}(max)c^{2}=~2\nu_{max}^{'}2m_{e}c^{2}\mu_{e}^{'}B~~,~~
\end{equation}
where the relation of $2\mu_{e}^{'}B_{\rm cr}/m_{e}c^{2}= 1$ is used.
Inserting Eq.(6) into Eq.(7) gives
\begin{eqnarray}
 \label{8}
&&p_{\bot}^{2}(max)c^{2} =~2\times \frac{1}{2B^{*}}(\frac{E_{\rm F}(e)}{m_{e}
c^{2}})^{2}\times 2m_{e}c^{2}\mu_{e}^{'}B \nonumber\\
&&\simeq B_{\rm cr}\times(\frac{E_{\rm F}(e)}{m_{e}c^{2}})^{2}\times
2m_{e}c^{2}\frac{e\hbar}{2m_{e}c} \nonumber\\
&&=\frac{m_{e}^{2}c^{3}}{e\hbar}\times(\frac{E_{\rm F}(e)}{m_{e}c^{2}})
^{2}\times 2m_{e}c^{2}\frac{e\hbar}{2m_{e}c}=~E_{\rm F}^{2}(e)~~.
\end{eqnarray}
In superhigh MFs, $E_{\rm F}(e)$ is determined by
\begin{equation}
\label{9}
E_{\rm F}(e)\simeq 43.44(\frac{B}{B_{\rm cr}})^{1/4}(\frac{\rho}{\rho_{0}}\frac{Y_e}{0.0535})^{\frac{1}{4}}~
~\rm MeV~~,~
\end{equation}
where $\rho_{0}= 2.8\times 10^{14}$~g~cm$^{3}$ is the standard nuclear
density (Gao et al. 2012b). Thus, we obtain
\begin{eqnarray}
\label{10}
&&p_{\bot}(max)= p_{\rm F}(e)\simeq \frac{E_{\rm F}(e)}{c}\nonumber\\
&&= 43.44\times(\frac{Y_{e}}{0.0535}\frac{\rho}{\rho_{0}}\frac{B}{B_{\rm cr}})
^{\frac{1}{4}}{\rm MeV/c} ~(B^{*} \geq 1~)~~.~~
\end{eqnarray}
As pointed out above, when $n= 0$, the electron Landau level is non-degenerate,
and $p_{z}$ has its maximum $p_{z}(max)$,
\begin{eqnarray}
\label{11}
&&p_{z}(max)= p_{\rm F}(e)\simeq \frac{E_{\rm F}(e)}{c}\nonumber\\
&&= 43.44\times(\frac{Y_{e}}{0.0535}\frac{\rho}{\rho_{0}}\frac{B}{B_{\rm cr}})
^{\frac{1}{4}}{\rm MeV/c}  ~(B^{*} \geq 1~)~~.~~
\end{eqnarray}
From Eq.(9) and Eq.(10), it's obvious that $p_{z}(max)=p_{\bot}(max)
= p_{\rm F}(e)$. The reason for this is
that in the interior of a magnetar, electrons are degenerate and super-relativistic,
and can be approximately treated as an ideal Fermi gas with equivalent pressures
in all directions, though the existence of Landau levels.  The equation of
$P_{e}$ in a superhigh magnetic field is consequently given by
\begin{eqnarray}
\label{12}
&&P_{e}= \frac{1}{3}\frac{2}{h^{3}}\int_{0}^{p_{\rm F}(e)}\frac{p_{e}^{2}c^{2}}
{(p_{e}^{2}c^{2}+ m_{e}^{2}c^{4})^{1/2}}4\pi p_{e}^{2}dp_{e}\nonumber\\
&&=1.412\times10^{25}\phi(x_{e}){\rm dynes~cm^{-2}}~
 \end{eqnarray}
where $\lambda_{e}=\frac{h}{m_{e}c}$ is the electron Compton wavelength,
$x_{e}=\frac{p_{\rm F}(e)}{m_{e}c}\simeq \frac{E_{\rm F}(e)}
{m_{e}c^2}=86.77\times(\frac{\rho}{\rho_{0}}\frac{B}{B_{\rm cr}}\frac{Y_{e}}
{0.0535})^{\frac{1}{4}}$), and $\phi(x_{e})$ is the polynomial $\phi(x_{e})=
\frac{1}{8\pi^2}[x_{e}(1+x_{e}^{2})^{\frac{1}{2}}
 (\frac{2x_{e}^{2}}{3}- 1)+ln[x_{e}+(1+x_{e}^{2})^{\frac{1}{2}}]]$.

When $\rho\geq 10^{7}$~g~cm$^{-3}$, $x_{e}\gg 1$, and $\phi(x_{e})
\rightarrow \frac{x_{e}^{4}}{12\pi^2}$. Thus, Eq.(11) can be rewritten as
\begin{equation}
\label{13}
P_{e}\simeq 6.266\times10^{30}(\frac{\rho}{\rho_{0}}\frac{B}{B_{\rm cr}}
\frac{Y_{e}}{0.0535})~~{\rm dyne~cm^{-2}}.
\end{equation}
It is obvious that $P_{e}$ increases sharply with increasing $B$ when the
values of $Y_{e}$ and $\rho$ are given.
\section{Magnetic effects on EoSs}
\subsection{Magnetic effects on the EoS of BPS model}
By introducing the lattice energy, Baym, Pethick \& Sutherland (1971)
(hereafter ``BPS model'') improved on Salpeter's treatment (Salpeter 1961), and described the nuclear
composition and EoS for catalyzed matter in complete thermodynamic
equilibrium below $\rho_{d}$. BPS model is one of most successful models describing matter of the outer crust.
According to BPS model, the matter energy density is given by
 \begin{equation}
 \label{14}
\varepsilon = n_{N}(W_{N}(A, Z) + \varepsilon_{L}(Z, n_{e})+
\varepsilon_{e}(n_e)~,~~
\end{equation}
where $n_{N}$ is the number density of nuclei, $W_{N}(A, Z)$ is the mass-
energy per nucleus (including the rest mass of $Z$ electrons and $A$ nucleons);
$\varepsilon_{e}$ is the free electron energy including the rest mass of
electrons in a unit volume; $\varepsilon_{L}$ is the $bcc$ Coulomb lattice energy
per nucleus,
\begin{equation}
\label{15}
 \varepsilon_{L}=-1.444Z^{2/3}e^{2}e^{2}n_{e}^{4/3}~~ , ~~
\end{equation}		
where the relations of $n_{N}=n_{B}/A$ and $n_{e}=Zn_{N}$ are used.
The matter pressure $p$ of the system is given by
\begin{equation}
\label{16}
 P= P_{e} + P_{L}= P_{e}+ \frac{1}{3}\varepsilon_{L}.
\end{equation}
 For a magnetic field $B^{*}\gg 1$, $P_{e}$ in Eq.(15) is given by
Eq.(12). Based on the above equations, we plot one schematic diagrams of QED effects on the
EOS of BPS model, as shown in Fig.1.
\begin{figure}[!htbp]
\centering
 \includegraphics[width=0.45\textwidth]{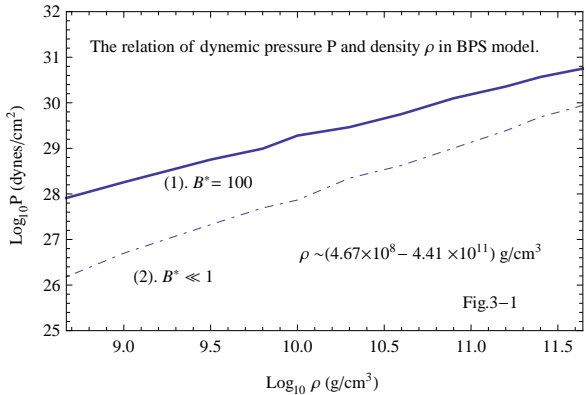}
\caption{$P$ vs. $\rho$ of BPS model below neutron drip.}
\label{fig_1}
\end{figure}
\subsection{The QED effects on the EOS of ideal $npe$ gas}
 We consider a homogenous ideal $npe$ gas under $\beta$-
equilibrium, and adopt ST83 approximation (Shapiro \& Teukolskysee 1983)
corresponding to the weak-field limit as the main method to treat EoS of this system
in the density range of $0.5\sim 2.5\rho_{0}$ where electrons are relativistic,
neutrons and protons are non-relativistic. When neutron pressure dominates, $\rho
\approx m_{n}n_{n}$, then $n_{n}=1.7\times 10^{38}(\frac{\rho}
{\rho_{0}})$~cm$^{-3}$ (Shapiro \& Teukolskysee 1983); employing $\beta$-
equilibrium and charge neutrality gives $n_{p}= n_{e}= 9.6\times
10^{35}(\frac{\rho}{\rho_{0}})^{2}$~cm$^{-3}$; $\beta$-equilibrium implies
energy conservation and momentum conservation ($p_{\rm F}(p)=p_{\rm F}
(e)$), we get $E_{\rm F}(e)= \mu_{n}= E_{\rm F}^{'}(n)
=p_{\rm F}^{2}(n)/2m_{n}=60(\frac{\rho}{\rho_{0}})^{2/3}$~MeV, and
$\mu_{p}=E_{\rm F}^{'}(p)=p_{\rm F}^{2}(p)/2m_{p}= 1.9
(\frac{\rho}{\rho_{0}})^{4/3}$~MeV; the isotropic matter pressure $P$ is given by
\begin{eqnarray}
\label{17}
&&P= P_{e}+ P_{p}+ P_{n}\nonumber\\
&&= \frac{m_{e}c^{2}}{\lambda_{e}^{3}}\phi(x_{e})+ \frac{m_{p}c^{2}}{\lambda_{p}^{3}}\phi(x_{p})+
\frac{m_{n}c^{2}}{\lambda_{n}^{3}}\phi(x_{n})~~, ~~
 \end{eqnarray}
where $x_{p}=~\frac{p_{\rm F}(p)}{m_{p}c}=
\frac{\sqrt{2m_{p}\mu_{p}}}{m_{p}c}~=~\sqrt{\frac{2\mu_{p}}{m_{p}c^2}}$, the
expression of $\phi(x_{p})$ is completely similar to that of $\phi(x_{\rm n})$.

Based on the above results, we gain the following useful formulae:
\begin{eqnarray}
\label{18}
&&P_{p}=1.169\times 10^{30}(\frac{\rho}{\rho_{0}})^{\frac{10}{3}}~~{\rm dynes~cm^{-2}}~~,\nonumber\\
&&P_{e}=1.825\times 10^{31}(\frac{\rho}{\rho_{0}})^{\frac{8}{3}}~~{\rm dynes~cm^{-2}}~~,\nonumber\\
&&P_{n}=6.807\times 10^{33}(\frac{\rho}{\rho_{0}})^{\frac{5}{3}}~~{\rm dynes~cm^{-2}}~~,\nonumber\\
&& Y_{e}=\frac{n_{e}}{n_{p}+n_{n}}\simeq \frac{n_{e}}{n_{n}}= 0.005647(\frac{\rho}{\rho_{0}})~~.~~
 \end{eqnarray}
Our methods to treat EOS of an ideal $npe$ gas (system) under $\beta$-equilibrium
in superhigh MFs are introduced as follows: Combining Eq.(9) with momentum
conservation  gives the chemical potential $\mu_{p}= E_{\rm F}
^{'}(p)= 1.005(\frac{B}{B_{\rm cr}}\frac{\rho}{\rho_{0}}\frac{Y_{e}}{0.0535})^{\frac{1}{2}}$~
MeV, and the non-dimensional variable $x_{p}= \sqrt{\frac{2\mu_{p}}{m_{p}c^2}}\simeq 4.626
\times 10^{-2}(\frac{B}{B_{\rm cr}}\frac{\rho}{\rho_{0}}\frac{Y_{e}}{0.0535})^{\frac{1}{4}}$;
Then we get
\begin{eqnarray}
\label{19}
 &&x_{n}= \sqrt{\frac{1}{m_{n}c^2}}(2\times(43.44(\frac{B}{B_{\rm cr}}\frac{\rho}{\rho_{0}}\frac{Y_{e}}{0.0535})^{1/4}~\nonumber\\
 &&-1.29 + 1.005(\frac{B}{B_{\rm cr}}\frac{\rho}{\rho_{0}}\frac{Y_{e}}{0.0535})^{1/2}))^{1/2}.
 \end{eqnarray}
The $\beta$-equilibrium condition gives the expression for
the isotropic matter pressure $P$,
\begin{eqnarray}
\label{20}
&&P= \frac{m_{e}c^{2}}{\lambda_{e}^{3}}\phi(x_{e})+ \frac{m_{p}c^{2}}{\lambda_{p}^{3}}\phi(x_{p})+
\frac{m_{n}c^{2}}{\lambda_{n}^{3}}\phi(x_{n})\nonumber\\
&&= 6.266\times 10^{30}(\frac{\rho}{\rho_{0}}\frac{B}{B_{\rm cr}}\frac{Y_{e}}{0.0535})+2.324\times 10^{26}\nonumber\\
&&\times(\frac{\rho}{\rho_{0}}\frac{B}{B_{\rm cr}}\frac{Y_{e}}{0.0535})^{\frac{5}{4}}+ 1.624\times 10^{38}\nonumber\\
&&\times\frac{1}{15\pi^2}(x_{n}^{5}-\frac{5}{14}x_{n}^{7}+\frac{5}{24}x_{n}^{9})~{\rm dyne~cm^{-2}}~~,
 \end{eqnarray}
where $x_{n}$ is determined by Eq.(19). The above equation always
approximately hold in an ideal $npe$
gas when $B^{*}\gg 1$ and $\rho~\sim 0.5\rho_{0}-~2\rho_{0}$. Based on the above equations,
we plot two schematic diagrams of QED effects on EOS of this $npe$
gas, as shown in Figs.2-3.
\begin{figure}[!htbp]
\centering
 \includegraphics[width=0.45\textwidth]{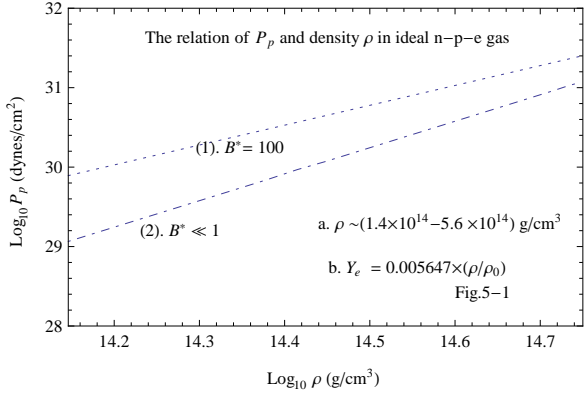}
\caption{$P_{p}$ vs. $\rho$ for an ideal $npe$ gas in a
superhigh MF.}
\label{fig_3}
\end{figure}
\begin{figure}[!htbp]
\centering
 \includegraphics[width=0.45\textwidth]{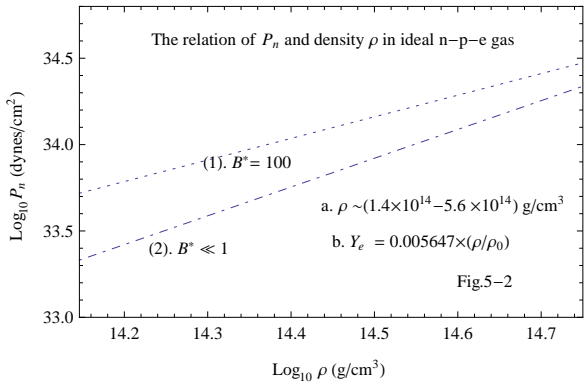}
\caption{$P_{n}$ vs. $\rho$ for an ideal $npe$ gas in a
superhigh MF.}
\label{fig_4}
\end{figure}
Both $P_{p}$ and $P_{n}$ increase
obviously with $\rho$ and $B$ for an ideal $npe$ gas.

\subsection{The QED effects on the total matter pressure and total energy density}
As discussed above, the pressures of fermions increase with $B$, the total matter pressure
increases with $B$. Due to a positive co-relation between the total energy density $\varepsilon$ and the total matter pressure, $\varepsilon$ also increases with $B$.

The stable configurations of a NS can be obtained from the well-known hydrostatic
equilibrium equations of Tolman, Oppenheimer and Volkov (TOV) for the pressure $P(r)$
and the enclosed mass $m(r)$,
\begin{eqnarray}
\label{21}
 &&\frac{dP(r)}{dr}=-\frac{G(m(r)+ 4\pi r^{3}P(r)/c^{2})(\rho+P(r)/c^{2})}{r(r-2Gm(r)/c^{2})}\nonumber\\
&&\frac{dm(r)}{dr} = 4\pi \rho r^{2}~,~~
\end{eqnarray}
where $G$ is the gravitational constant. For
a chosen central value of $\rho$, the numerical integration of Eq.(21)
provides the mass-radius relation.  In
Eq.(21), the pressure $P(r)$ is the gravitational collapse pressure, and
always be balanced by the total dynamics pressure, $P$; the central density
$\rho$ is proportional to the matter energy density $\epsilon$; the
enclosed mass, $m(r)$, increases with the central density $\rho$ when
$r$ is given.

As we know, the magnetic effects can give rise to an anisotropy of
the total pressure of the system to become anisotropic (Bocquet et al.
1995, Paulucci et al. 2011). The total energy momentum tensor due to both matter
and magnetic field is to be given by
\begin{equation}
\label{22}
T^{\mu \nu}=  T^{\mu \nu}_{m}+ T^{\mu \nu}_{B},
\end{equation}
where,
\begin{equation}
\label{23}
T^{\mu \nu}_{m}= \epsilon_{m}u^{\mu}u^{\nu}- P_{m}(g^{\mu \nu}- u^{\mu}u^{\nu}),
\end{equation}
and
\begin{equation}
\label{24}
T^{\mu \nu}_{B}= \frac{B^{2}}{4\pi}(u^{\mu}u^{\nu}- \frac{1}{2}g^{\mu \nu})- \frac{B^{\mu}B^{\nu}}{4\pi}~.~
\end{equation}
The first term in Eq.(24) is
equivalent to magnetic pressure, while the second
term causes the magnetic tension.
Due to an excess negative pressure or tension along the direction to the
magnetic field, the component of $T^{\mu \nu}_{B}$ along the field, $T^{zz}
_{B}$, is negative. Thus, the total pressure in the parallel direction to MF
can be written as
\begin{equation}
\label{25}
P_{\|}= P_{m}- \frac{B^{2}}{8\pi},
\end{equation}
and that perpendicular to MF, $P_{\bot}$, is written as
\begin{equation}
\label{26}
P_{\bot}= P_{m}+ \frac{B^{2}}{8\pi}-{\cal M}B~,~
\end{equation}
where ${\cal M}$ is the magnetization of the system, and ${\cal M}B$ is the
magnetization pressure (Perez et al. 2008, Ferrer et. al. 2010). In this work magnetars universally have
typical dipole MFs $\sim(10^{14}-10^{15})$ G and inner field strengths not more than $10^{17}$ G, under which the system magnetic moment satisfies
${\cal M} < B$, a condition that can be justified for any medium that is not ferromagnetic, the effect of AMMs of
nucleons on the EOS are insignificant and thus ignored (Ferrer et. al. 2015).
It's obvious that the total pressure of the system becomes anisotropic, that is $P_{\bot}> P_{\|}$, which could lead
to the Earth-like oblatening effect.

 According to our calculations, when $B^{*}= 100 $, $P_{m}\sim
10^{33}-10^{34}$~dynes~cm$^{-2}$ and $\frac{B^{2}}{8\pi}\sim 10^{29}-10^{30}$. Hence, in
this presentation, we consider that the component of the total energy
momentum tensor along the symmetry axis becomes positive, $T^{zz}> 0$, since the total
matter pressure increases more rapidly than the magnetic pressure.
We propose that the component of the total energy momentum tensor along the
symmetry axis becomes positive, since $P_m$ always grows more rapidly than the magnetic
pressure. The magnetic tension along the direction to the magnetic field will be
responsible for deforming a magnetar along MF, and turns the
star into a kind of oblate spheroid. Be note that such a deformation in shape
might even render a more compact magnetar endowed with canonical strong surface
fields $B\sim 10^{14-15}$~G.  Also, such a deformed magnetar could have a more
massive mass because of the positive contribution of the magnetic field energy
to EOSs of a magnetar.
\section{To modify $P_e$ in superhigh MFs }
According to atomic physics physics, the higher the orbit quantum
number $l$ is, the larger the probability of an electron's transition
(this transition is referred to the transition from a higher energy
level into a lower energy level) is.  Analogous to atomic energy level, in a superhigh MF, the
easier an electron's transition from a higher Landau level into a lower Landau level,
Thus, the higher the electron Landau level number $n$, the lower the stability of
the Landau level. Owing to the uncertainties of microscopic states, we introduce a new quantity, $g_{n}$, the stability coefficient of
electron Landau level in a superhigh MF, and assume that $g_{n}$ decreases with $n$ as an exponential form,
\begin{equation}
\label{27}
g_{n} = g_{0}n^{\alpha},
\end{equation}
where $g_{0}$ is the stability coefficient of the ground-state Landau level of electrons, $\alpha$ is the
Landau level stability index, and is restricted to be $\alpha<0$. From Eq.(27), it is obvious that $g(n)$ is a function of $n$ and
$\alpha$, and the higher $n$ is, the smaller $g_{n}$ is (except for $g_{1}=g_0$).

According to the Pauli exclusion principle, electron energy state number in a unit volume, $N_{pha}$,
should be equal to electron number in a unit volume, $n_{e}$. Considering the electron Landau
level stability coefficient $g_{n}$, and summing
over electron energy states in a 6-dimension phase space, we can
express $N_{pha}$ as follows:
\begin{eqnarray}
\label{28}
 &&N_{pha}= n_{e}=N_{A}\rho Y_{e} \nonumber\\
 &&=\frac{2\pi}{h^{3}}\int dp_{z}\sum_{n = 0}^{n_{m}
 (p_z,\sigma,B^{*})}\sum g_{n}\nonumber\\
 &&\times\int \delta(\frac{p_{\perp}}{m_{e}c}-[(2n+1+\sigma)B^{*}]
 ^{\frac{1}{2}}) p_{\perp}dp_{\perp},
 \end{eqnarray}
where $N_{A}$ is the Avogadro constant. When $n_{m} \geq$ 6, the summation
formula can be approximately replaced by the following integral equation
\begin{equation}
 \label{29}
\sum_{n = 0}^{n_{m}}n^{\alpha+\frac{1}{2}}\simeq \int_{0}^{n_{m}}n^{\alpha+\frac{1}{2}}dn
 = \frac{2}{2\alpha+ 3}n^{\alpha+\frac{3}{2}}_{m}.
\end{equation}
Thus, Eq.(28) can be rewritten as
\begin{eqnarray}
\label{30}
 &&N_{pha} = N_{A}\rho Y_{e}=\frac{2^{\frac{7}{2}}}{2\alpha+3}\pi \sqrt{B^{*}}
 (\frac{m_{e}c}{h})^{3}g_{0}\nonumber\\
 &&\int_{0}^{\frac{p_{F}}{m_{e}c}} [(\frac{E_{F}(e)}{m_{e}c^{2}})^{2}
  -1-(\frac{p_{z}}{m_{e}c})^{2}]^{\alpha+\frac{3}{2}}d(\frac{p_{z}}{m_{e}c}).
 \end{eqnarray}
 After a complicate deduction process, we get an non-dimensional momentum
\begin{equation}
\label{31}
x_{e}=\frac{p_{F}(e)}{m_{e}c}=C\left[\frac{Y_{e}}{0.05} \frac{\rho}{\rho_{0}}\right]
^{\frac{1}{2(\alpha+2)}}(B^{*})^{\frac{\alpha+1}{2(\alpha+2)}},
\end{equation}
where $C$ is a constant, which is determined by
\begin{eqnarray}
\label{32}
&&C=(\frac{0.05\rho_{0}N_{A}(2\alpha+3)}{2^{2(1-\alpha)}\pi g_{0}I(\alpha)})^{\frac{1}{2(\alpha+2)}}(\frac{h}{m_{e}c})^{\frac{3}{2(\alpha+2)}}\nonumber\\
&&\simeq (337.12)^{\frac{3}{2(\alpha+2)}}(\frac{2\alpha +3}{2^{2(1-\alpha)} g_{0}I(\alpha)})^{\frac{1}{2(\alpha+2)}}.
\end{eqnarray}
with $\int_{0}^{1}(1-t^{2})^{3/2+\alpha}$, and $t=p_{z}c/E_{F}(e)$. If $\alpha$ and $g_{0}$ are determined,
the expressions of $E_{F}(e)$ and $P_{e}$ in superhigh MFs  will be modified accordingly.
To exactly determine the values of $\alpha$ and $g_{0}$ is an interesting and important task,
but is beyond of this paper. Since this is an original work in which some
uncertainties could exist, to further modify and perfect our model
should be considered in our future studies, especially to further investigate QED effects on the
EoSs using an improved expression of $P_e$ in a superhigh MF.
\section{ Conclusions}
In this presentation, we derived a general expression for
electron pressure, which holds in a superhigh MF,
considered QED effects on EoSs of neutron star matter, and discussed an anisotropy of the
total pressure in superhigh MFs. Compared with a common pulsar, a magnetar could be a more
compact oblate spheroid-like deformed NS, due to the anisotropic total pressure;
an increase in the maximum mass of a magnetar is expected because of the positive
contribution of the magnetic field energy to EoS.
\acknowledgements
This work was supported by Xinjiang Natural Science Foundation No.2013211A053.
This work is also supported in part by Chinese National Science Foundation through grants
No. 11273051, 11173041, 11133001, 11447165, 11173042 and 11373006, National Basic Research Program of China grants 973 Programs 2012CB821801,
the Strategic Priority Research Program ``The Emergence of Cosmological Structures'' of Chinese Academy of Sciences
through No.XDB09000000, and by a research fund from the Qinglan project of Jiangsu Province.
%\newpage11173041,11173042, 11273051,


\begin{thebibliography}{}
%\bibitem{} Author1, A.B., Author2, C.D.: 2001, AN 322, 1
\bibitem{} Baym, G., Pethick, C., Sutherland, P.: 1971, ApJ 170, 29
 \bibitem{}Bocquet, M., et al.,: 1995, A\&A 301, 757
 \bibitem{}Canuto, C., Ventura, J.: 1977, Fund. Cosmic Phys. 2, 203
 \bibitem{}Du, J., Luo, Z.Q., Zhang, J.: 2014, Ap\&SS, 351, 625
 \bibitem{}Du, Y.J., et al.: 2009, MNRAS 399, 1587
 \bibitem{}Franzon, B., Dexheimer, V., Schramm, S.,: 2015, arXiv:1508.04431 (submtted)
  \bibitem{}Ferrer, E. J., Incera, V. de la., Keuth, J. P.,et al.: 2010, Phys. Rev. C. 82, 065802
  \bibitem{}Ferrer, E. J., Incera, V. de la., Paret, D. M.,et al.: 2015, Phys. Rev. D. 91, 085041
\bibitem{} Gao, Z.F., Wang, N., Yuan, J.P., et al.: 2011a, Ap\&SS 332,129
  \bibitem{} Gao, Z.F., Wang, N., Yuan, J.P., et~al.: 2011b, Ap\&SS 333, 427
  \bibitem{} Gao, Z.F., Wang, N., Song, D.L., et al.: 2011c, Ap\&SS 334, 281
  \bibitem{} Gao, Z.F., Peng, Q.H., Wang, N., et al.: 2011d, Ap\&SS 336, 427
  \bibitem{} Gao, Z.F., et~al.: 2012a, Chin. Phy. B. 21(5), 057109
  \bibitem{} Gao, Z.F., et~al.: 2012b, Ap\&SS 342, 55
  \bibitem{} Gao, Z.F., et~al.: 2013, Mod. Phys. Lett. A. 28(36), 1350138
  \bibitem{} Gao, Z.F. et al.: 2014, Astron. Nachr. 335, No.6/7, 653
  \bibitem{} Gao, Z.F. et al.: 2015, MNRAS, arXiv:1505.07013 (accepted)
   \bibitem{}Lai, D., Shapiro, S. L.:  1991, ApJ 383, 745
   \bibitem{}Lai, X. Y., et al.: 2013, MNRAS 431, 3290
  \bibitem{} Liu J.-J.: 2012, Chin.Phys.lett. 29, 122301
  \bibitem{}Liu J.-J.: 2013, MNRAS 433, 1108
  \bibitem{}Liu J.-J.: 2014, MNRAS 438, 930
  \bibitem{}Liu J.-J.: 2015, Ap\&SS 357, 93
  \bibitem{} Paulucci, L.,: 2011, Phys. Rev. D. 83(4), 043009
 \bibitem{}P\'{e}rez Mart\'{i}nez, A., et al.: 2008, Int. J. Mod. Phys. D 17, 210
  \bibitem{} Salpeter, E. E.: 1961, ApJ 134, 669
 \bibitem{}Shapiro, S. L., Teukolsky, S. A.: 1983, ``Black Holes, White Drarfs, and Neutron Stars'',  New
    York, Wiley-Interscience
  \bibitem{} Thompson, C., Duncan, R.C.: 1996, ApJ 473, 322
  \bibitem{}Xu, Y., et al.: 2013, Chin. Phys. Lett. 29, 059701
\end{thebibliography}
\end{document}